# Spatio-temporal isolation of attosecond soft X-ray pulses in the water window

Francisco Silva[1], Stephan Teichmann[1], Seth L. Cousin[1], Michael Hemmer[1], Jens Biegert[1,2,*]

[1]*ICFO-Institut de Ciencies Fotoniques, Mediterranean Technology Park, 08860 Castelldefels (Barcelona), Spain*

[2]*ICREA-Institució Catalana de Recerca i Estudis Avançats, 08010 Barcelona, Spain*

We demonstrate experimentally the isolation of single attosecond pulses at the carbon K-shell edge in the soft-X-ray water window. Attosecond pulses at photon energies that cover the principal absorption edges of the building blocks of materials are a prerequisite for time resolved probing of the triggering events leading to electronic dynamics such as exciton formation and annihilation. Herewith, we demonstrate successful isolation of individual attosecond pulses at the carbon K edge (284 eV) with a pulse duration below 400 as and with a bandwidth supporting a 30 as pulse duration. Our approach is based on spatio-temporal isolation of ponderomotively shifted harmonics and validates a straightforward and scalable approach for robust and reproducible attosecond pulse isolation.

The power of attosecond science lies in its capability for time resolving electronic motion on its native ultrafast timescale [1, 2], i.e. observation of the previously intractable "triggering events" leading to electronic excitation, formation of charge density waves, changes of chemical bonds or the emergence of ultrafast magnetism, just to mention a few examples. The fundamental requirements for such investigations are isolated attosecond-duration pulses in synchronicity with excitation or interrogation pulses and at photon energies that permit probing the relevant mechanisms. For instance, the low efficiency of organic solar cells is related to a lack of knowledge of the ultrafast events within the energy transformation processes, i.e. ultrafast exciton formation and annihilation. Modern organic solar cells consists of metal ligand hydrocarbon complexes and attosecond duration pulses with photon energy of 284 eV, as demonstrated here, will permit probing exciton dynamics at the K-shell of the carbon atom of the solar cell.

Isolated attosecond pulses are generated by confining high harmonic generation (HHG) driven with an intense laser to a short temporal window that is on the order of the field cycle duration of the driving laser. This condition can be met by either producing few-cycle duration pulses [3] or through a variety of methods that effectively confine the recollision to a similar time span [4-6]. In order to generate repeatable attosecond pulses and to prevent random time shifts which would range on the order of the field cycle duration, control over the generating laser's electric carrier to envelope (CEP) field waveform is required. The maximum reachable photon energy from HHG, the so called cutoff, can be determined with the classical three step model of recollision [7] which is dependent on the harmonic generation medium, the laser peak intensity and scales with the square of the laser wavelength. The demanding requirements for single attosecond pulse generation have so far only been met with advanced Ti:Sapphire laser sources at 800 nm, which have confined attosecond generation largely to photon energies below 150 eV. Ponderomotive scaling of HHG to keV photon energies has been demonstrated by increasing the Ti:Sapphire laser's peak intensity [8] but the attainable non-phase-matched harmonic yield is low and further intensity scaling is limited by depletion of the medium's ground state. Exploiting wavelength scaling is therefore the route most actively pursued by many groups in the field with a keV cutoff being demonstrated when driven by a mid-IR source [9]. This has resulted in an upsurge in development of long wavelength ultrafast light sources based on parametric processes. One standing issue for HHG was the unfavorable scaling of the single atom response with laser wavelength [10, 11]. This was shown however to be largely mitigable, by employing high pressure phase-matching [12] which results in appreciable photon yields [13-15]. Recently, our group demonstrated the first soft X-ray absorption fine structure (XAFS) measurement at the carbon K-edge from a solid material

[15] thereby validating the feasibility of this approach for real applications in solid state physics.

In order to produce isolated attosecond pulses at high photon energies, high pressure phase-matching of long wavelength driven HHG may employ ionisation gating [5, 18], limiting emission to a single laser cycle [19]. Nevertheless, to be useful for time resolved measurements, an additional requirement is driving attosecond generation with a CEP controlled source since the absence of CEP stability would result in random timing slips of the attosecond emission on the order of the half cycle duration of the driving field; a 2000 nm driver wavelength corresponds to 3000 as for the duration of half a cycle. Recently, the first CEP dependent HHG spectra were demonstrated at 300 eV [13] and at 400 eV [15]. The to date unsolved problem is the isolation of a single attosecond pulse in a photon energy range above 150 eV. We address this crucial issue by demonstrating experimentally at the carbon K-edge at 284 eV spatio-temporal isolation (STI) of a single attosecond pulse in the water window.

**Implementation of wavefront rotation for soft-X-ray emission at the carbon K-edge.** Our laser system generates pulses at 1850 nm wavelength at a repetition rate of 1 kHz with a pulse duration of 13 fs. These pulses carry an energy of 230 µJ and are CEP stable to 80 mrad rms over 1h [15]. We use a 100 mm focal length spherical mirror to focus the pulses into the HHG target which consists of a 1.5 mm outer diameter tube with 0.3 mm entrance and exit holes through which the beam is focused; see the Methods section for further details and Fig. 1 for a schematic.

The laser radiation is filtered out by transmitting the harmonic beam through a 100 nm free-standing aluminium filter, whereafter the harmonics are refocused with a grazing incidence ellipsoidal mirror into a home built soft-X-ray spectrograph for analysis. The spectrograph consists of an aberration corrected 2400 l/mm flat field grating (Hitachi) and a cooled back-illuminated X-ray charge-coupled device camera (PIXIS-XO-2048B, Princeton Instruments). For STI of the attosecond emission, we employ a wavefront rotation scheme in the generation medium, also termed attosecond lighthouse [16, 17] or photonics streaking [18]. Wavefront rotation at 1.3 mrad/fs is implemented by inserting a thin (uncoated) fused silica wedge with a 5 degree apex angle before the focusing mirror. The maximum reachable peak intensity on target is 0.35 PW/cm$^2$ which is reduced to 0.3 PW/cm$^2$ when inserting the uncoated wedge. The wavefront rotation of the generating field translates directly into wavefront rotation of the generated harmonic field, leading to a directional change of the emission. Figure 1 illustrates the effect with a measured spatio-spectral and simulated spatio-temporal distribution of the generating pulse.

**Photonic streaking of 300 eV soft-X-ray emission.** To experimentally demonstrate our STI approach for isolation of a single attosecond pulse in the water window, we use neon as a generation medium as it facilitates HHG at the carbon K-edge at 284 eV with the pulse intensities available from our setup and for 3.6 bar backing pressure; the pressure dependence was recently demonstrated in Ref. [15]. The short pulse duration, and consequentially strong dependence on the CEP, permits an unambiguous investigation of the feasibility of isolating attosecond structures at this unprecedentedly high photon energy range using photonic streaking as each successive attosecond burst is launched in a different direction (here called beamlet), and thus is readily distinguishable. The measurement reported in Fig. 2, performed without WFR shows that as the CEP is changed, we observe a continuous movement of a single beamlet with fairly constant angular extent over the entire CEP range. The clear dependence of the angular drift with CEP, and the clean angular distribution is already an indication of emission of an attosecond pulse structure [18] since even a small pre- or post- pulse content (on the order of a few percent) would manifest itself in modulation of the angular content; see the inset in Fig. 2, right for a comparison.

**Spatio-temporal isolation of a single attosecond pulse in the water window.** We investigate the angular separation of beamlets and therefore our ability to steer and isolate single attosecond structures in the water window. Inserting the wedge reduces the peak intensity on target by 15% but still results in a 300 eV cutoff; this value is also inferred from the classical cutoff law [21]. Figure 3 shows the result of the measurement for three different relative CEP values of $\Phi$, $\Phi+\pi/3$ and $\Phi+2\pi/3$ with the spectrograph imaging the focal plane of the ellipsoidal mirror along the energy axis. Shown to the right is the emerging spectrum for angular integration over 2.5 mrad and for each case.

Figure 3(a) exhibits a fairly broad spectrum supporting a 59 as duration pulse. Next, we increment the CEP by $\pi/3$ which results in a broad and structureless spectral emission shown in Fig. 3(b) and thus supports the shortest 57 as duration water window pulse. Further incrementing CEP to $2\pi/3$, we observe in Fig. 3(c) a slight downshift of the highest emitted photon energy by 10 eV to 290 eV whereas the overall spectral shape remains. The persistence of the spectra for the isolated emission and the angular emission dependence on CEP is the first experimental evidence of isolated attosecond pulse generation in the water window at the carbon K-edge. Based on this clear demonstration of STI and the generation of single attosecond structures, we turn to a numerical code (described in Methods) to simulate the wavefront rotation effect in order to gain further insight into the interplay of the various quantum trajectories and the effectiveness of STI

Shown in Fig. 4(a) is the simulated spatio-temporal profile of the generating pulse and the effect of wavefront rotation. The far-field angular-spectral profile is depicted in (b) and shows the expected interferences due to emission from the different trajectories (long and short). The corresponding spectrogram is shown in (c). Our measurements (Fig. 3) do not exhibit signatures of such trajectory interferences and are in accordance with the findings in Ref. [16] which point at contributions from only the short trajectories. Thus we numerically removed the long trajectory contribution in our single atom calculations; the results are shown in the bottom row of Fig. 4. The far field pattern of Fig. 4(e) shows qualitative agreement with the measurement in Fig. 3. Figure 4(b,d) shows how the half cycle emissions are streaked onto the CCD camera. We find that the most prominent half cycle at -1 fs is emitted on axis, and is registered at the camera around 0 mrad. The spectrogram of the resulting soft-X-ray emission is shown in Fig. 4(f) and demonstrates prominent emission at the half cycle cutoff. The next, less prominent, half cycle emission at 2 fs is slightly upshifted and detected at 3 mrad. For completeness, the case for only the long trajectories is included in the Supplementary Information but does not show the same level of agreement with the measurement as Fig. 4(e). For comparison we also show harmonic emission for the case without wavefront rotation over the identical CEP range (Fig 5). Finally, we find that the simple single atom model qualitatively matches with our experimental investigations and evidences a predominant contribution from the short trajectory. The strong dependence of the cutoff and angular emission on CEP, as well as the clean isolation of single attosecond bursts, are first experimental evidence of isolated attosecond pulse generation in the water window at the carbon K-edge of 284 eV and show the applicability of STI to unprecedentedly high photon energies.

**Efficiency of STI.** Next we establish the efficiency of the spatio-temporal isolation method for water window radiation and its influence on photon yield and pre- and post- pulse contrast. For practical purposes, we contrast the implemented attosecond pulse isolation scheme against generation with a few-cycle pulse without wavefront rotation - see Fig. 5.

We find that STI leads to a five fold decrease in peak photon yield (blue curve Fig. 5(a)) compared to the implementation without STI (black curve). For a comparison of the flux for the attosecond emission however, the case is favorable for STI. In order to compare the two cases, we rely on our simulations. With STI, an iris can simply be inserted into the soft-X-ray beam to isolate a single attosecond pulse structure; the resulting photon yield is shown for the blue curve in Fig. 5(a). Without STI, the 100 nm Al filter will not sufficiently suppress emission from pre- and post- half cycles, thereby resulting in multiple attosecond pulse emission. Increasing the thickness of the soft-X-ray filter to 400 nm Al achieves a pre/post pulse contrast of 1:5 and calculated soft-X-ray pulse duration of 230 as (Fig. 5(b)). In contrast, STI results in a superior 1:20 contrast but a longer calculated 355 as duration (Fig. 5(d)). For comparison, we also show the performance of a state of the art multilayer mirror [22] for the water window (green curve Fig. 5(a)). By contrasting the relative efficiency of all schemes for generating a single attosecond pulse we find that the STI scheme presents a very attractive solution.

**Conclusions.** We have demonstrated isolation of a single attosecond pulse at the carbon K-shell edge at 284 eV in the water window using spatio-temporal isolation based on wavefront rotation. Our results validate the approach as a viable route for the isolation of a single attosecond pulse in a soft-X-ray energy regime in which the commonly used spectral filtering scheme with metal foils remains to be demonstrated, or whenever tunability of the emission energy may be required. Additionally, tuning emission to two attosecond bursts which are emitted into two different angular directions presents a route for the implementation of efficient soft X-ray pump / soft X-ray probe schemes in the high photon energy range by spatial recombination of two successive attosecond bursts and their relative pump probe delay via independent refocusing optics. Another attractive aspect of the spatial isolation method is the marginal penalty in photon flux, which, when compared to spectral filtering in the same energy regime, is even energetically superior. Our experimental demonstration is the first proof of the feasibility of implementation of such a single attosecond pulse selection regime for ponderomotively scaled sources which can reach the water window spectral regime and may become a major asset for the evolution of attosecond science into the energy regime important for material science and biology.

Methods

**Experiment.** High-energy pulses from a Ti:Sapphire amplifier system with 40 fs duration, 7 mJ energy at 1 kHz repetition rate are frequency converted to 1.85 um using an optical parametric amplifier, resulting in CEP-stable 0.8 mJ pulses with 40 fs duration. These pulses are spectrally broadened through nonlinear propagation in a hollow core fibre filled with 1.5 bar of argon, which both broadens the spectrum and introduces adequate dispersion for subsequent compression down to 13 fs with bulk material [15]. The CEP-stable pulses are then focused with an f=100 mm spherical mirror into a gas target (1.5 mm outer diameter, 0.5 mm inner diameter) with 0.3 mm diameter entrance and exit holes, filled with neon at a backing pressure of 3.6 atm. The residual fundamental radiation is filtered with a 100 nm free standing aluminium foil and the transmitted radiation is refocused using a grazing-incidence ellipsoidal mirror (Zeiss). A homebuilt spectrograph, consisting of a motorised slit, flat-field imaging reflective grating (Hitachi, 2400 lines/mm) and a cooled, back-illuminated CCD (Princeton Instruments) is used to measure the spatio-spectral profile of the radiation in function of CEP. Photon numbers are extracted using the measured spectra, the camera quantum efficiency and the measured grating diffraction efficiency.

**Simulations.** The simulation results presented in the discussion section were obtained with a model based on two different HHG codes, one for the 3D propagation without WFR, the other for the microscopic response with WFR. In the case of no WFR, cylindrical symmetry is assumed and nonlinear propagation using pseudo-spectral methods is used to numerically propagate the driving laser electric field. Dispersion, diffraction, self-phase modulation, plasma dephasing and absorption are taken into account [23]. Ionisation rates are calculated using the Ammosov-Delone-Krainov (ADK) formula [24]. After calculation of the driving laser field the single-atom HHG response is computed at every point of the propagation grid (resampled to provide appropriate resolution) using an augmented Lewenstein model [11, 25] with an electron birth factor that incorporates the ADK ionisation rate [26-28] and photorecombination cross section [29] for the recombination step [30]. Gaussian input pulses in the spatial and temporal domains were considered, and the target was considered to have a density of $1.2 \times 10^{19}$ atoms/cm$^3$. The simulated target thickness was the full inner diameter (0.5 mm). This simulation was used to calculate the results in figures 5b and c. In the case of WFR, the single atom response was calculated across the spatial dimension of the beam, using the code described above. The WFR was introduced by modelling the electric field using the measured quantities in Figs 1d and 1e, and using Gaussian integrated spectral and spatial profiles to minimise numerical noise. This simulation was used to calculate the results of Figs. 4, 5d and 5e. The pulse is propagated into the far field through fourier analysis, in order to calculate the spatio-spectral profile of Fig. 4b.

**Acknowledgements**  We acknowledge support from MINISTERIO DE ECONOMA Y COMPETITIVIDAD through Plan Nacional (FIS2011-30465-C02-01), the Catalan Agencia de Gestió D'Ajuts Universitaris i de Recerca (AGAUR) with SGR 2014-2016, Fundacio Cellex Barcelona, and funding from LASERLAB-EUROPE, grant agreement 228334.

**Author contributions**  F.S., S.T., S.L.C. acquired the experimental data. J. B. conceived the attosecond setup. M. H., S.L.C. and J.B. designed the laser system. J.B. and F.S. wrote the manuscript.

**Competing Interests**  The authors declare that they have no competing financial interests.

**Correspondence**  Correspondence and requests for materials should be addressed to Jens Biegert at jens.biegert@icfo.eu

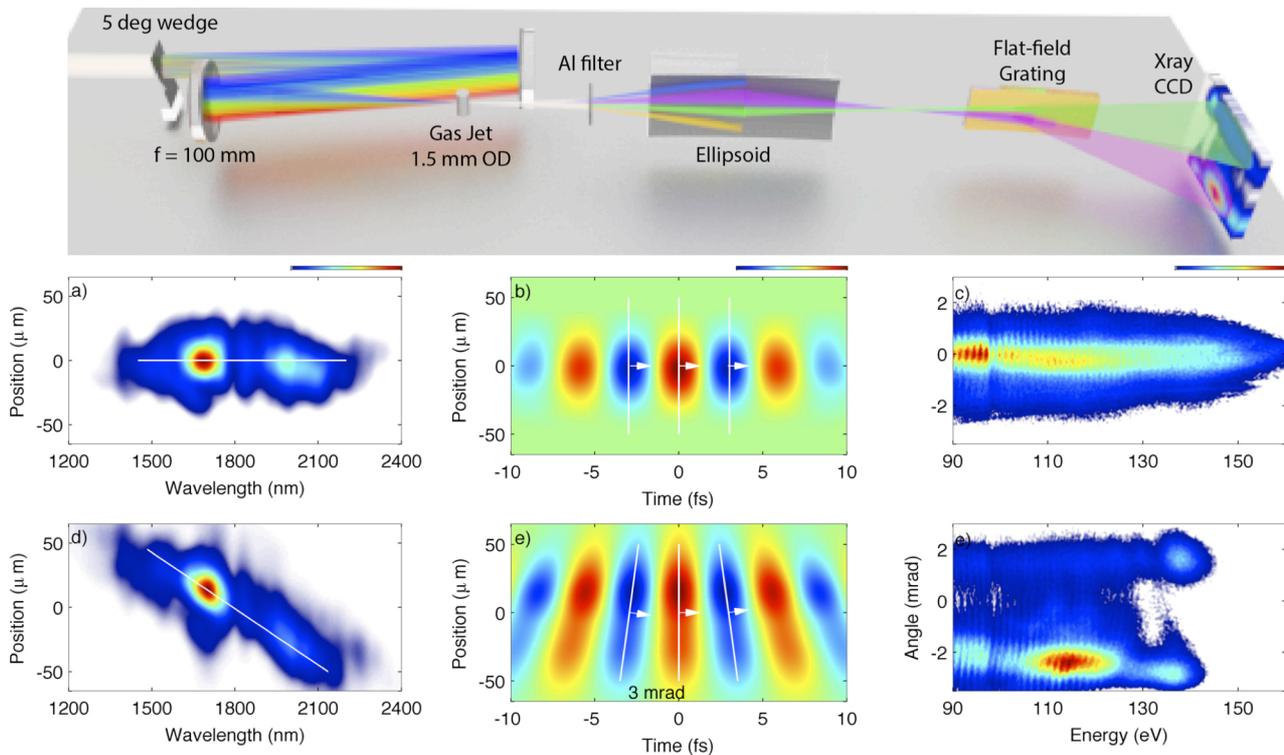

**Figure 1: Experimental Setup and Conditions.** Top: Schematic of the experimental Setup. (a) Spatio-spectral profile in the gas jet focus without WFR. (b) Corresponding calculated electric field. (c) Measured HHG spatio-spectral profile generated in 1 bar Argon without WFR. (d) Spatio-spectral profile in the gas jet focus with WFR. (e) Corresponding calculated electric field. (f) Measured HHG spatio-spectral profile generated in 1 bar Argon with WFR, revealing a 3 mrad separation between beamlets.

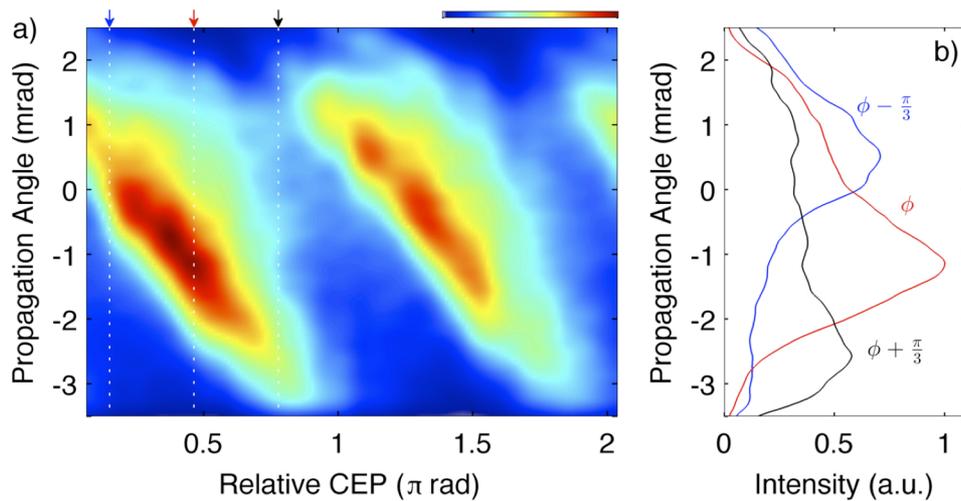

**Figure 2: CEP control of emission angle (a)** Spatial profile of the emitted radiation in function of CEP, revealing a clear control of the emission angle. **(b)** Spatial profile for three CEPs.

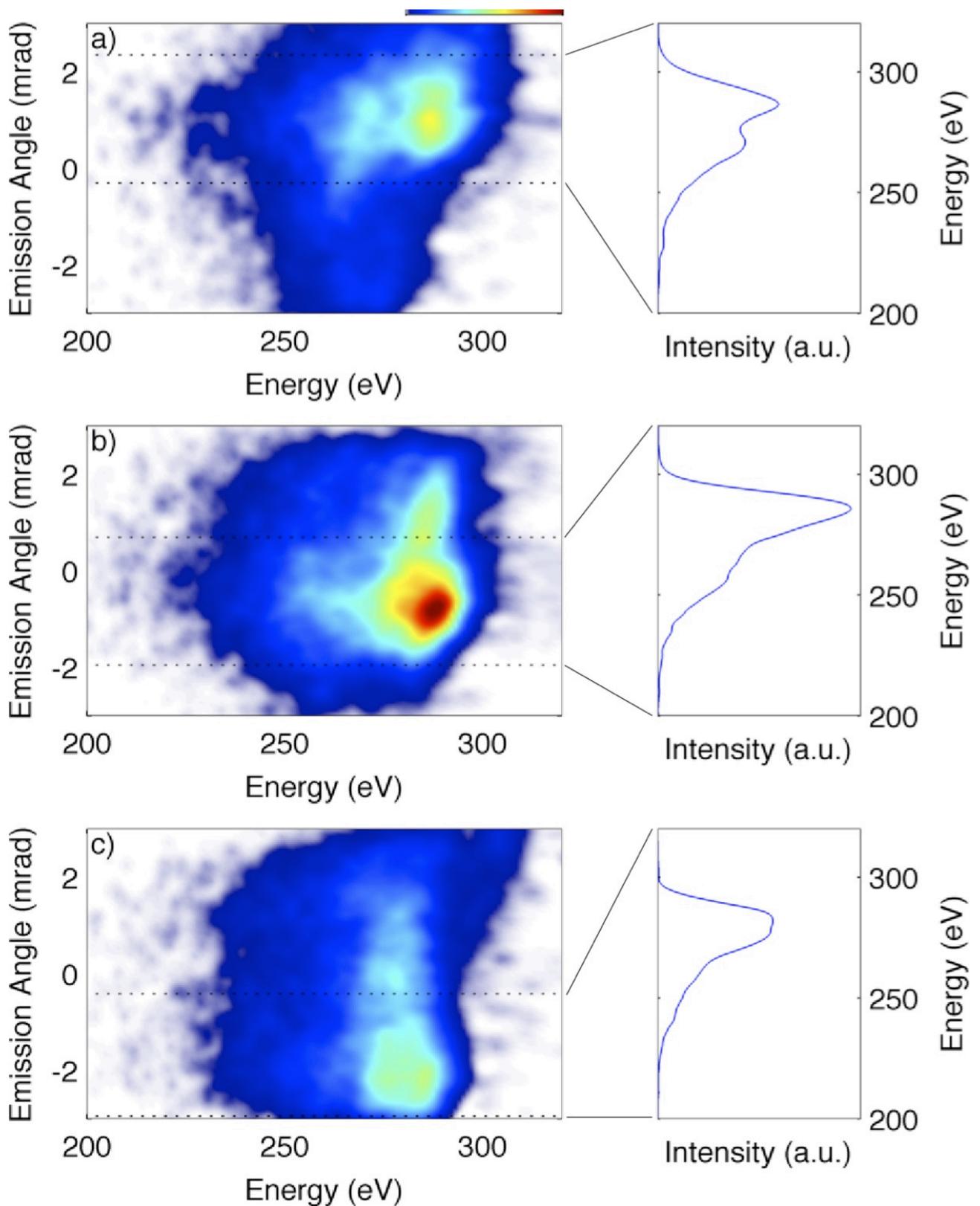

Figure 3: Isolated spatio-temporal profiles for three different emission angles (a-c). Dotted lines: 2.5 mrad integration region for the spectra on the right.

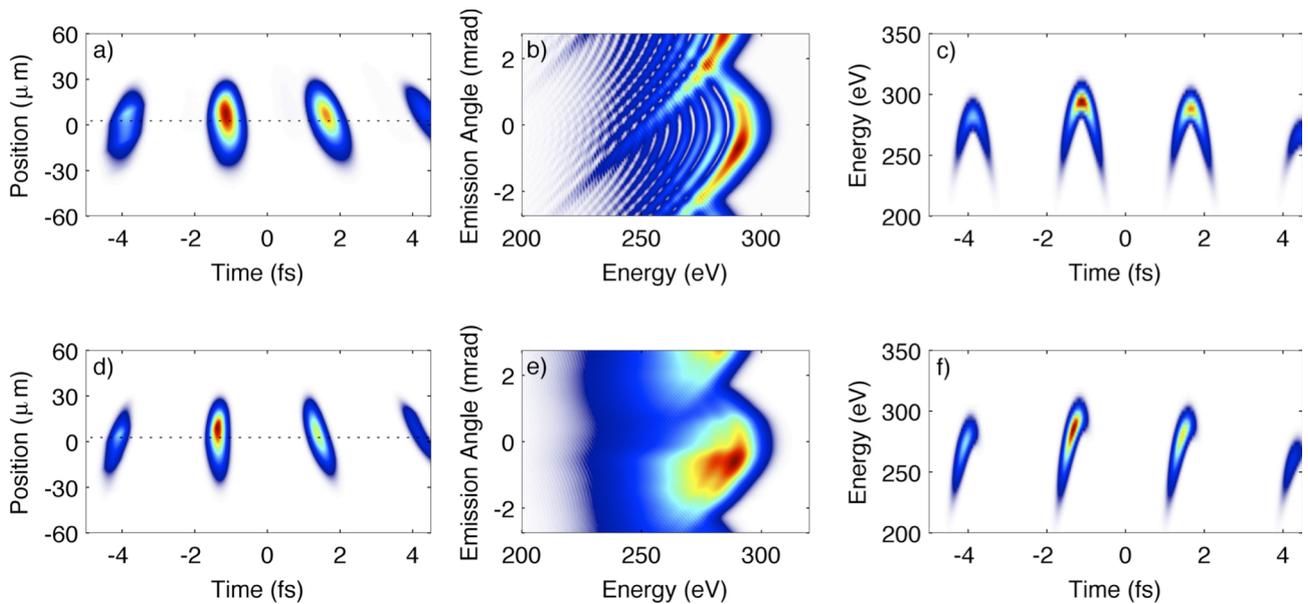

**Figure 4: Single-atom response WFR simulation results.** Shown in the top row (a-c) are results including short and long trajectories, whereas the bottom row (d-f) shows results for identical simulation values, but with the long trajectory numerically removed. (a, d) show the generated spatio-temporal profile in focus; (b,e) angular-spectral profile in the far-field, for an angular range of 5.5 mrad; (c,f) spectrogram of the pulse, calculated for the electric field along the dotted line in Fig. 4(a). The case for long trajectories is only displayed in the SI for completeness.

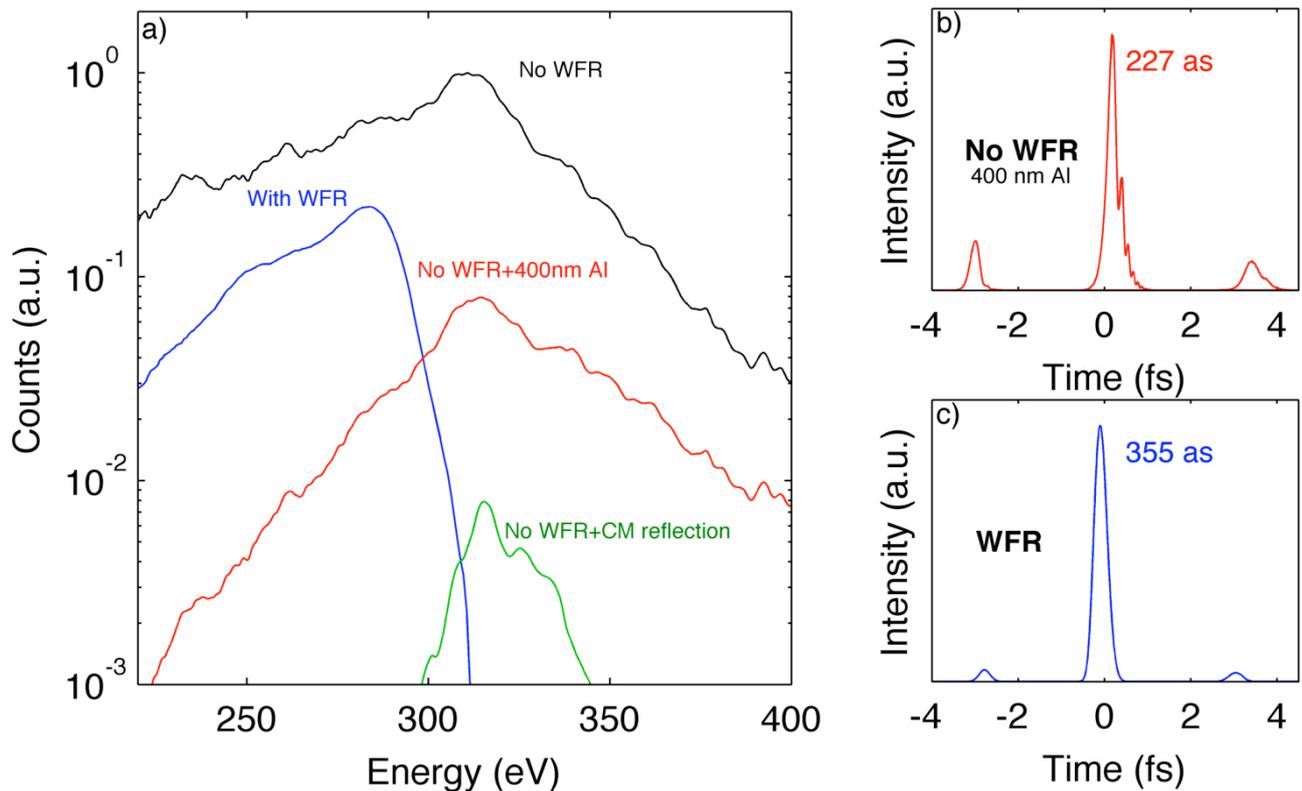

**Figure 5: Photon yield and calculated pulse profiles. (a)** Comparison of photon yields on target. Chirped mirror reflection profile from Ref. 22. **(b).** Simulated pulse profile in the case without wavefront rotation, after transmission through a 400 nm Al filter, leading to a pre/post pulse contrast of 1:5. **(c)** Simulated pulse profile in the case of wavefront rotation, leading to an increased pre/post pulse contrast of 1:20.